\documentclass[12pt,english]{article}
\usepackage{times}
\usepackage[T1]{fontenc}
\usepackage[latin1]{inputenc}
\usepackage{geometry}
\usepackage{setspace}
\makeatletter

\usepackage{babel}
\makeatother
\begin{document}

\title{Reaction diffusion processes on random and scale-free networks}

\author{Subhasis Banerjee, Shrestha Basu Mallik$^{*}$ and Indrani Bose \\
Department of Physics, Bose Institute, 93/1, A. P. C. Road, Kolkata-700009\\
{*} Department of Physics, Indian Institute of Technology, Kanpur,
India.}

\maketitle
\begin{abstract}
We study the discrete Gierer-Meinhardt model of reaction-diffusion
on three different types of networks: regular, random and scale-free.
The model dynamics lead to the formation of stationary Turing patterns
in the steady state in certain parameter regions. Some general features
of the patterns are studied through numerical simulation. The results
for the random and scale-free networks show a marked difference from
those in the case of the regular network. The difference may be ascribed
to the small world character of the first two types of networks.

\end{abstract}

\section{Introduction}

Reaction-diffusion (RD) processes provide the basis for pattern formation
in several physical, chemical and biological systems {[}1-4{]}. One
of the most prominent examples of such processes is based on the Turing
mechanism {[}4{]}. In a celebrated paper, Turing {[}5{]} showed that
a diffusion-driven instability may occur when infinitesimal perturbations
are applied to an initially homogeneous system of reacting and diffusing
chemicals. The instability gives rise to spatially heterogeneous stationary
patterns in the steady state. This is illustrated by considering a
system of two chemicals : the activator and the inhibitor. The activator
is autocatalytic, i.e., it promotes its own production as well as
that of the inhibitor. The inhibitor, as the name implies, inhibits
the production of the activator. The diffusion coefficient of the
inhibitor is moreover much larger than that of the activator. Consider
a homogeneous distribution of the activator and the inhibitor in the
RD system. Perturbation is applied to the system through small local
increases in the activator concentration. This gives rise to further
increases in the concentration of the activator in the local regions
due to autocatalysis. The inhibitor concentration is also enhanced
locally. The inhibitor, with a higher diffusion coefficient, reaches
the surrounding regions first and prevents the activator from spreading
into these regions. A nonequilibrium steady state is obtained if the
decay of the activator and the inhibitor is offset by a constant supply
of the chemicals. This state is characterised by a stationary distribution
of islands of high activator concentration in a sea of high inhibitor
concentration. The islands constitute what is known as the Turing
pattern.

The Gierer-Meinhardt (GM) model provides a mathematical description
of the RD processes, leading to Turing instability, through the partial
differential equations

\begin{center}\[
\frac{\partial a}{\partial t}=D_{a}\bigtriangledown^{2}a+\rho_{a}\frac{a^{2}}{h}-\mu_{a}a\]
\end{center}

\[
\,\,\,\,\,\,\,\,\,\,\,\,\,\,\,\,\,\,\,\,\,\,\,\,\,\,\frac{\partial h}{\partial t}=D_{h}\bigtriangledown^{2}h+\rho_{h}a-\mu_{h}h\,\,\,\,\,\,\,\,\,\,\,\,\,\,\,\,\,\,\,\,\,\,\,\,(1)\]
 where `$a$' and `$h$' are the concentrations of the activator and
the inhibitor, $D_{a}$, $D_{h}$ are the respective diffusion coefficients,
$\mu_{a}$,~$\mu_{h}$ the removal rates and $\rho_{a}$,~$\rho_{h}$
the cross-reaction coefficients. The RD processes described by Eq.(1)
are defined in the continuum. In this paper, we study the formation
of Turing patterns in networks with a discrete structure. Three types
of network are considered : regular, random and scale-free. The RD
processes are described by a simple discretization of Eq.(1). Such
discretization, necessary for obtaining numerical solutions of partial
differential equations, provides a  coupled map model for the networks.
Section 2 contains a description of the coupled map model the dynamics
of which, in specific parameter regions, lead to the formation of
Turing patterns in the steady state of the networks. Section 3 contains
concluding remarks.

\section{Turing patterns in RD networks }

Consider a network of \textit{N} nodes connected to each other via
links. At each node \textit{i} (\textit{i}=1,............,\textit{N}),
the concentration of the activator and the inhibitor are given by
$a_{i}$ and $h_{i}$. Time \textit{t} is discretized in steps of
unity and the evolution of the concentration variables is described
by the coupled map equations\\

\[
a_{i}(t+1)=a_{i}(t)+D_{a}\sum_{j}C_{ij}(a_{j}(t)-a_{i}(t))+\rho_{a}\frac{a_{i}^{2}(t)}{h_{i}(t)}-\mu_{a}a_{i}(t)\,\,\,\,\,\,\,\,\,\]

\[
\,\,\,\,\,\,\,\,\,\,\,\, h_{i}(t+1)=h_{i}(t)+D_{h}\sum_{j}C_{ij}(h_{j}(t)-h_{i}(t))+\rho_{h}a_{i}^{2}(t)-\mu_{h}h_{i}(t)\,\,\,\,\,\,\,\,\,\,\,\,\,\,\,\,\,\,(2)\]
The coupling matrix \textit{C} is symmetric with diagonal elements
zero and $C_{ij}$ =1 if the nodes \textit{i} and \textit{j} are connected
and is zero otherwise. The diffusive coupling in Eq.(2) has the form
of a finite difference approximation. Under the same approximation,
Eq.(1) reduces to Eq.(2) with $\delta t=1$, $(\delta x)^{2}=1,$
where $\delta t$ is the time increment and $\delta x$ the mesh size.
In the case of networks, however, Eq.(2) defines the coupled map model.

Three different types of networks have been considered : regular,
random and scale-free, each with 2500 nodes. The regular network is
a square lattice for which the degree of each node, given by the number
of links associated with the node, is exactly four. The degree distributions
of the random and the scale-free networks are Poissonian and power-law
respectively {[}6,7{]}. The random network is described by the Erd\"os-R\'{e}nyi
(ER) model {[}8{]}. The network has \textit{N} nodes and each pair
of nodes is connected with probability $p_{ER}$ so that the total
number of links in the network is $n=p_{ER}\, N(N-1)/2$. The scale-free
network is generated following the prescription of Barab\'{a}si et
al {[}9{]}. One starts with a small number $m_{0}$ of nodes. In every
time step, a new node with $m\leq m_{0}$ links is added (m=2 in the
present simulation). The new node is connected to an existing node
\textit{i} with probability $\prod(k_{i})$ which depends on the degree
\textit{k$_{\textrm{i}}$} of the node \textit{i}. The preferential
attachment probability is given by 

\begin{center}\[
\,\,\,\,\,\,\,\,\,\,\,\,\,\,\,\,\,\,\,\,\,\,\,\,\,\prod(k_{i})=\frac{(k_{i}+1)}{\sum_{j}(k_{j}+1)}\,\,\,\,\,\,\,\,\,\,\,\,\,\,\,\,\,\,\,\,\,\,\,\,\,\,\,\,(3)\]
 \end{center}

After \textit{T} time steps, the network has \textit{T+}$m_{0}$ nodes
and $mT+m_{0}(m_{0}-1)/2$ links assuming all $m_{0}$ initial sites
to be connected. The evolution rule leads to a scale-free network
when the network size is significantly large. The average number of
links per node is fixed to be four in both the random and scale-free
networks in order that a meaningful comparison with the square lattice
results can be made. With $\rho_{a}=\mu_{a}$ and $\rho_{h}=\mu_{h}$,
the steady state $(a_{i}(t+1)=a_{i}(t),\,\, h_{i}(t+1)=h_{i}(t))$
is given by a homogeneous distribution of activator and inhibitor
concentration at all the nodes, say, $(a_{i},h_{i})=(1,1)$ for all
\textit{i}=1,.......\textit{N}. This homogeneous steady state is taken
to be the initial state of each network. The steady state is perturbed
by small amounts $(a_{i}\rightarrow1+\delta a_{i}\,,h_{i}\rightarrow1+\delta h_{i}\,)$
at each node using a random number generator. The amount of perturbation
is chosen to be the same in the cases of the activator and the inhibitor,
i.e., $\delta a_{i}=\delta h_{i}$. Time evolution of the perturbed
system is determined with the help of Eq.(2). The homogeneous steady
state is stable if the perturbed system returns to it after some time
steps. The full parameter space corresponding to Eq.(2) contains a
region in which the homogeneous steady state is stable. There is another
region in the parameter space corresponding to which the perturbed
system exhibits Turing instability. The steady state to which the
instability leads is obtained if $a_{i}$ and $h_{i}$ change by less
than $10^{-4}$ on five consecutive iterations of Eq.(2) for all \textit{i}.
The state is characterised by a stationary pattern of Turing peaks
corresponding to gradients of activator concentration in local regions.
The height of a peak is defined as the magnitude of the concentration
variable at the highest point of the gradient. Fig. 1 shows a distribution
of Turing peaks in the steady state of a regular network (square lattice
with 2500 nodes) with parameter values given by $D_{a}=0.00055,$~$D_{h}=0.01,$$\,\,\rho_{a}=\mu_{a}=0.00055,\,\,\rho_{h}=\mu_{h}=0.0011$.

We now describe the main results of our study on some general aspects
of Turing patterns in the steady states of the regular, random and
scale-free networks. The initial random number seed is taken to be
the same in each case so that the pattern of perturbations at the
nodes is identical. Our first observation relates to the fact that
the formation of Turing pattern is most favourable, i.e., occurs over
a wider region in parameter space in the case of the regular network.
Figs. 2-6 are obtained by varying the diffusion coefficient $D_{a}$
of the activator and keeping the diffusion coefficient $D_{h}$ of
the inhibitor fixed. The variable along the \textit{x}-axis in each
case is the ratio $d=D_{h}/D_{a}$. The values of $\mu_{a}=\rho_{a}$
and $\mu_{h}=\rho_{h}$ are kept fixed at $\mu{}_{a}=0.00055$ and
$\mu_{h}=0.0011$. Fig. 2 shows the average concentration of the activator
versus \textit{d} for the regular, random and scale-free networks.
The average is found to be the highest in the case of the regular
network. Fig. 3 shows the number of nodes $N_{a}$ at which the activator
concentration $a_{i}$ is greater than or equal to a threshold value,
say 2, versus \textit{d}. The number of such nodes appears to be the
largest in the case of the regular network over almost the full range
of \textit{d}. For smaller values of \textit{d}, the number of nodes
with $a_{i}\geq2$ in the case of the scale-free network is greater
than the corresponding number for the random network but for larger
values of \textit{d}, the numbers are more or less equal. Fig. 4 shows
the maximum peak height (equivalently, the highest value of the activator
concentration at a node) versus \textit{d} for the three networks.
The maximum height attained in the steady state is more or less the
same in the cases of the random and scale-free networks. This value
is greater than that in the case of a regular network over the full
range of \textit{d}. Each data point in Figs. 2-4 is an average over
five realizations of the steady state. The different realizations
are obtained by changing the initial seed of the random number generator.
Fig. 5 shows the connectivity of the node \textit{i} at which the
activator concentration has the highest value in the steady state,
versus \textit{d}. The plot shows an interesting plateau structure.
The connectivity is found to shift to higher values for larger \textit{d}.
Fig. 6 shows the same plot for the scale-free network. The plateau
structure is similar to that observed in the case of the random network.

From the results obtained, the major conclusion one arrives at is
that for the general features of Turing patterns described above,
the regular network is markedly different from the random and scale-free
networks. The latter two types of networks have more or less similar
features. The variation of pattern related quantities as a function
of \textit{d} is smoother in the case of the regular network whereas
in the cases of the other two types of networks, the variation is
much less smooth. We now look for possible explanations of the results
obtained. The important differences between the three networks is
that both the random and the scale-free network have a small world
(SW) character, i.e., any pair of randomly chosen nodes is connected
by a path consisting of a small number of links. This is not true
in the case of a regular network like the square lattice. In the first
two cases, one can define an average path length $L_{av}$, which
is the average of the shortest paths connecting all pairs of nodes
in the network, the length of a path being given by the number of
links contained in the path. In the case of the random network, $L_{av}\sim$log\textit{N}
where \textit{N} is the number of nodes in the network. Scale-free
networks, with degree exponents in the range 2<$\gamma$<3 are ultra
-small, i.e., $L_{av}\sim$loglog\textit{N} {[}10{]}. In the case
of a regular network, $L{}_{av}$ scales with some power of \textit{N},
rather than log\textit{N}. Effective communication between the nodes
is thus much greater in the cases of random and scale-free networks.
For RD processes, one can define two length scales, namely the activator
and the inhibitor decay lengths, given by $l_{a}=\sqrt{(D_{a}/\mu_{a})}$
and $l_{h}=\sqrt{(D_{h}/\mu_{h})}$ respectively. The decay length
provides an estimate of the distance over which molecules diffuse
before disappearing due to decay. Turing instability requires short
range activation and long range inhibition, i.e., $l_{a}<l_{h}$.
In obtaining Fig. 2-6, the parameters $D_{h},\,\mu_{a},\,\mu_{h}$
are kept constant and $D_{a}$ is varied over the range in which Turing
instability occurs. Thus the decay length $l_{h}$ of the inhibitor
is constant with the value $l_{h}=3.01$ and the decay length of the
activator is decreased from the value $l_{a}=1.0$ towards zero. Since
the number of nodes in the networks studied is 2500, the average path
length $L_{av}$ in the cases of random and scale-free networks is
of similar magnitude as $l_{h}$. In obtaining the data points in
Figs. 2-6, the same random number seeds are chosen for each realization
(each data point is an average over five realizations), so that the
only variation comes from changing the diffusion coefficient $D_{a}$
of the activator.

Increase in activator concentration at a node is possible if there
is a net inflow of activator from the other nodes. Since the largest
magnitude of the decay length $l_{a}$ of the activator is 1.0 ( the
first data point), increase in activator concentration through diffusion
is minimal. Increase in activator concentration at a node occurs mainly
through autocatalysis. For this, it is desirable that the amount of
activator diffusing away from the  node in question is small. Higher
concentration of activator is obtained if there is a net outflow of
inhibitor from the node. The steady state concentration of activator
at a node is a balance between autocatalysis, inhibition and decay.
The average concentration of activator, over the range of $d=D_{h}/D_{a}$
studied, is the highest in the case of the regular network. In the
cases of random and scale-free networks, the SW feature leads to a
greater overall amount of inhibition so that the number of nodes at
which the steady state concentration of the activator is above a threshold
value (2 in our case) as well as the average concentration of the
activator are lower (Figs. 2 and 3). The maximum value of the activator
concentration in the steady state, when concentrations at all the
nodes are considered, is, however, found to be higher than that in
the case of a regular network (Fig. 4). The maximum value increases
as $D_{a}$ is lowered . In the case of the regular network, the increase
occurs over a small range of values of \textit{d} and then the maximum
concentration value saturates. The increase is steeper and over a
much wider range in the cases of random and scale-free networks. In
general, there are some nodes at which the activator concentration
in the steady state is significantly higher than the maximum value
of the concentration in the case of a regular network. However, a
larger number of nodes with activator concentration above a threshold
value gives rise to a higher average concentration in the case of
a regular network. The nodes in the random and scale-free networks
have variable connectivity. At some of these nodes, high activator
concentration is obtained in the steady state due to autocatalysis
being more dominant over inhibition than in the case of a regular
network. One example of this is illustrated in Figs. 5 and 6. The
plots represent connectivity $p_{highest}$ of the node, at which
the highest activator concentration occurs, versus $d=D_{h}/D_{a}$
with $D_{h}$ kept constant. The connectivity of a node is the number
of links to which the node belongs.The plot exhibits an interesting
plateau structure. In this case, the data points are not averaged
over five realizations as averaging is expected to destroy the plateau
structure. The magnitude of $p_{highest}$ is found to increase on
reducing $D_{a}$, i.e., increasing $d=D_{h}/D_{a}$ ($D_{h}$ constant).
As the same pattern of perturbations is applied for obtaining each
data point, the only difference in each case arises from the changed
magnitude of $D_{a}$. Initially, $p_{highest}$ is low, around 2.
A small number of links implies greater isolation and consequent enhancement
of autocatalysis which is a local effect. We now discuss the origin
of a plateau. As $D_{a}$ is decreased, the amount of activator remaining
at the node and participating in autocatalysis increases. The amount
of inhibitor reaching or leaving the node is the same as $D_{h}$
remains constant. As a result, the highest activator concentration
in the steady state increases with lowered values of $D_{a}$ (Fig.
4). The plateau occurs as long as the node, at which the highest activator
concentration is obtained, remains unchanged. The plateau ends when
it becomes advantageous for $p_{highest}$ to be raised. Increased
number of links may not be detrimental as, because of a lowered value
of $D_{a}$, the net amount of activator diffusing away is still small
whereas an increased number of links allows a greater amount of inhibitor
to leave the node, leading to higher activator concentration in the
steady state. The plateau structure is seen for other sets of parameter
values also. Fig. 7 shows the distribution of activator concentration
in the steady states of the regular and scale-free networks for $D_{a}=0.000015$
and $D_{h}=0.01$ . The bin size is taken to be 0.5. An average over
ten realizations has been taken. The data for the random network are
not plotted for clarity. The distribution in this case is similar
to that of the scale-free network.

\section{Summary and Discussion }

In this paper, we have studied the formation of Turing patterns in
the cases of regular, random and scale-free networks. The RD processes
are described by a model which is a simple discretized version of
the GM model. Formation of Turing patterns is most favourable in the
case of the regular network (square lattice). The size of the network
is kept the same in each case. The average degree of nodes is four
in the cases of the random and scale-free networks so that a meaningful
comparison with square lattice results can be made. Some general features
of Turing patterns in the steady state have been studied like the
average activator concentration versus $d=D_{h}/D_{a}$, the ratio
of the diffusion coefficients of the inhibitor and the activator,
the number of nodes $N_{a}$ at which the activator concentration
$a_{i}$ is greater than or equal to a threshold value versus \textit{d},
the highest value of $a_{i}$ versus \textit{d}, and the distribution
of activator concentration amongst the network sites. In each case,
the results for the random and the scale-free networks are markedly
different from thoose of the regular network. The differences can
be explained in terms of the small world character of the first two
types of networks, These networks also exhibit an interesting plateau
structure in a plot of the connectivity of the node \textit{i} at
which the activator concentration has the highest value in the steady
state, versus \textit{d}. Fig. 7 provides clear evidence that the
distribution of the activator concentration amongst the nodes of the
network is markedly different in the cases of regular and random/scale-free
networks. The scale-free network, considered in this paper, is of
the Barab\'{a}si - type with the degree exponent $\gamma\sim3$.
For this high value of $\gamma$, the number of highly connected nodes
is very small which may be the reason why the scale-free and random
networks exhibit similar features. The value of $D_{h}$, the diffusion
coefficient of the inhibitor has been kept constant to identify the
systematic trends associated with the variation of $D_{a}$. The range
of $D_{h}$ values for which Turing patterns form in the steady state
is not sufficiently long to study the variation with respect to $D_{h}$,
keeping $D_{a}$ fixed. We have, however, verified that the results
reported in this paper hold true for other values of $D_{h}$ as well
as for different parameter sets. 

RD processes are associated with many chemical and biological systems
{[}1-4{]}. In neurobiology, there is considerable evidence that synaptic
transmission may not be the exclusive mechanism for neurotransmission
in brain functions. There are suggestions {[}11{]} that nonsynaptic
diffusion neurotransmission plays a fundamental role in certain sustained
brain functions which include vigilance, hunger, mood, responses to
certain sensory stimuli as well as abnormal functions like mood disorder,
spinal shock, spasticity and drug addiction. Liang {[}11{]} has proposed
a RD neural network model to demonstrate the advantages of nonsynaptic
diffusion from a computational viewpoint. The RD processes considered
are described by the GM model. The spatial Laplacian operator is approximated
by finite differences since in a neural network, the neurons are located
at discrete positions. The activator and the inhibitor of the GM model
are produced by the neurons of the network. Due to Turing-type instabilities,
the network can support multiple simultaneous spatiotemporal organization
processes. In fact, the Turing islands of activator concentration
gradients may correspond to distinct areas of brain activity. Liang
considered the RD processes on a square network but a real neural
network is more like a random graph with a SW character {[}6,7{]}.
The present study clearly shows that the Turing patters on random
scale-free networks have characteristics distinct from those in the
case of a regular network. The existence of higher concentration gradients
in the first two cases may imply sharper signalling response whereas
lower numbers of Turing peaks possibly favour the emergence of nonoverlapping,
i.e., distinct functional areas. Biological networks like gene transcription
regulatory and metabolic reaction networks have a scale-free character
{[}6,7,12{]}. These networks serve as scaffolds for various RD processes.
There has been suggestions that specific biological activities may
be controlled by concentration gradients of appropriate type arising
out of Turing-like instabilities {[}1{]}. In general, RD processes
may give rise to other types of instabilities leading to stationary,
oscillatory and travelling wave patterns. It will be of considerable
interest to find specific examples of patterns generated by RD processes
in biological networks. A deeper question relates to the small world
character of such networks and its role in essential biological phenomena.

Acknowledgement: S. Basu Mallik carried out the work under the Summer
Student Programme of the Indian Academy of Sciences, Bangalore.

\begin{description}
\item [Fig.]1 Distribution of the Turing peaks in the steady state of a
square lattice with 2500 nodes for parameter values $D_{a}=0.00055$,
$D_{h}=0.01$, $\rho_{a}=\mu_{a}=0.00055$ and $\rho_{h}=\mu_{h}=0.0011$.
\item [Fig.]2 Average activator concentration versus $d=D_{h}/D_{a}$ in
the steady state for $D_{h}=0.01,$ $\rho_{a}=\mu_{a}=0.00055$ and
$\rho_{h}=\mu_{h}=0.0011.$ The data points, solid square, solid circle
and solid triangle correspond to regular (square-lattice), random
and scale-free networks respectively.
\item [Fig.]3 Number of nodes $N_{a}$ at which the activator concentration
is greater than equal to a threshold value, versus $d=D_{h}/D_{a}$
for $D_{h}=0.01$, $\rho_{a}=\mu_{a}=0.00055$, and $\rho_{h}=\mu_{h}=0.0011$.
The data points, solid square, solid circle and solid triangle corresponds
to regular (square-lattice), random and scale-free networks respectively.
\item [Fig.]4 Maximum activator concentration versus $d=D_{h}/D_{a}$ in
the steady state for $D_{h}=0.01,$ $\rho_{a}=\mu_{a}=0.00055$ and
$\rho_{h}=\mu_{h}=0.0011.$ The data points, solid square, solid circle
and solid triangle corresponds to regular (square-lattice), random
and scale-free networks respectively.
\item [Fig.]5 $p_{highest}$ versus $d=D_{h}/D_{a}$ in the steady state
for $D_{h}=0.01,$ $\rho_{a}=\mu_{a}=0.00055$ and $\rho_{h}=\mu_{h}=0.0011$
in the case of the random network.
\item [Fig.]6 $p_{highest}$ versus $d=D_{h}/D_{a}$ in the steady state
for $D_{h}=0.01,$ $\rho_{a}=\mu_{a}=0.00055$ and $\rho_{h}=\mu_{h}=0.0011$
in the case of the scale-free  network.
\item [Fig.]7 Distribution of activator concentration amongst the network
sites in the steady states of the regular (solid dots) and scale-free
(open circles) networks.\end{description}

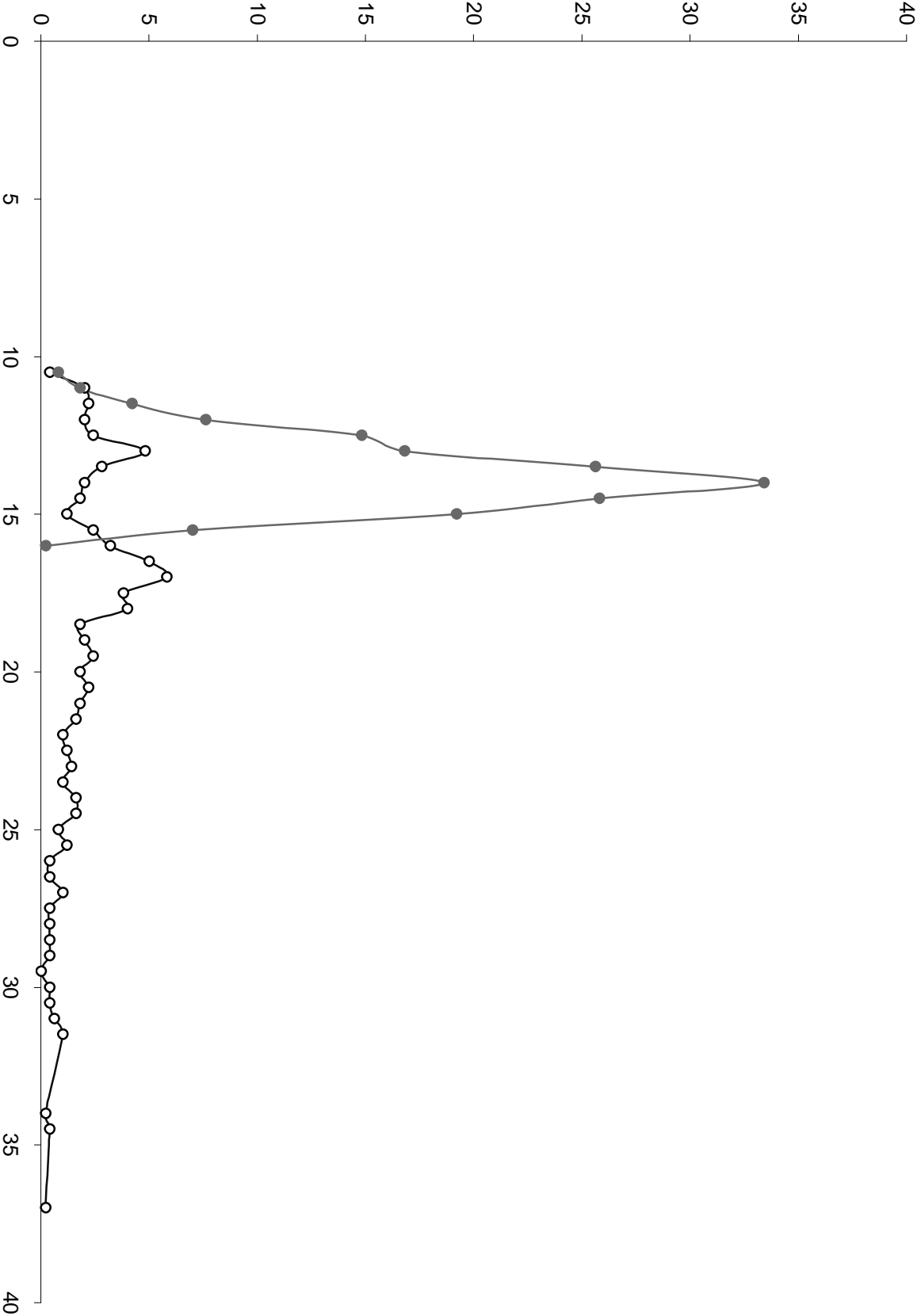
\end{document}